\documentclass[aps,pra,showpacs,superscriptaddress,floatfix,nofootinbib]{revtex4}

\usepackage{psfrag,graphicx} \usepackage{dcolumn}
\usepackage{amsmath,amssymb} \usepackage{bm}
\usepackage[dvips]{epsfig,color} \usepackage[english]{babel}

\newcommand{\la}{\langle} \newcommand{\ra}{\rangle}

\newcommand{\be}{\begin{equation}}
\newcommand{\bea}{\begin{eqnarray}}
\newcommand{\eea}{\end{eqnarray}}
\newcommand{\ee}{\end{equation}}

\def\tr{\mathrm{Tr}}

\begin{document}

\title{2D Multipartite Valence Bond States in Quantum Antiferromagnets}

\author{E. Rico} 
\affiliation{Institut fuer Theoretische Physik, Universitat Innsbruck, Technikerstrasse 25, A-6020 Innsbruck, Austria.}  
\affiliation{Institute of Quantum Optics and Quantum Information of the Austrian Academy of Science, A-6020 Innsbruck, Austria.} 
\affiliation{Fakultat fuer Physik, Universitat Wien, Boltzmanngasse 5, A-1090 Wien, Austria.} 
\author{H.J. Briegel} 
\affiliation{Institut fuer Theoretische Physik, Universitat Innsbruck, Technikerstrasse 25, A-6020 Innsbruck, Austria.}  
\affiliation{Institute of Quantum Optics and Quantum Information of the Austrian Academy of Science, A-6020 Innsbruck, Austria.} 

\date{\today}

\begin{abstract}
A quantum anti-ferromagnetic spin-1 model is characterised on a 2D lattice with the following requirements: i) The Hamiltonian is made out of nearest neighbour interactions. ii) It is homogeneous, translational and rotational invariant. iii) The ground state is a real singlet state of SU(2) (non-chiral). iv) It has a local spin-1 representation.

Along the way to characterise the system, connections with classical statistical mechanics and integrable models are explored. Finally, the relevance of the model in the physics of low dimensional anti-ferromagnetic Mott-Hubbard insulators is discussed. 
\end{abstract}

\maketitle 

\section{Introduction}

The description of the low temperature properties of a given physical system is one of the most fundamental problems in condensed matter physics. Following the renormalization group ideas the problem of analysing the thermodynamic features of a system can sometimes be recast into the characterisation of the infrared states that are compatible with the symmetries of the problem. Concepts like the order parameter and the related broken symmetries give us deep insights about the properties of the ground state and phase transition between different states. 

So far, universality classes like Fermi-Landau theory appear as paradigms for the classification of quantum systems. Nevertheless, as soon as the interactions between the constituents in the system become more and more important, the Fermi-Landau picture breaks down and new universality classes or infrared points are needed to study or describe the physics of the problem. For instance, quantum states like the BCS \cite{Bardeen:1957mv} or the Laughlin \cite{Laughlin:1983fy} wave functions are known to describe the basic features of conventional superconductors or fractional quantum hall systems, respectively.

In the area of quantum magnetism, the Neel state and the spin wave theory appear as the classic paradigm (see for example \cite{Fradkin:1991ho,Auerbach:1994yp,Essler:2005ly} and references therein). In the last decades, a new family of states known as spin liquids \cite{Rokhsar:1988uq,Moessner:2001kx} has attracted a lot of attention (see also \cite{Misguich:2005fk} and references therein). One of the reasons for such attention is due to the fact that a subclass of these spin liquid states, the valence bond states (\emph{VBS}), has been purposed as the right picture to describe the low energy properties of Mott-Hubbard anti-ferromagnetic insulators, which have a deep relation with the high temperature superconductivity in cuprates \cite{Anderson:1987oq,Zhang:1988nx} and colossal magneto-resistance in manganites \cite{Ramirez:1997fk}.

From the quantum information community, ideas and tools \cite{Bennett:2000at,Nielsen:2000ne,:2001fz,Esteve:2003uo} have brought new insights and a fresh perspective to problems in quantum physics, in general, and highly correlated quantum systems \cite{Amico:2007fk}, in particular. In fact, it was pointed out by Preskill\cite{Preskill:1999he} that this interdisciplinary area of research could provide a better understanding of questions that appear in many-body quantum entanglement and more refine classifications of different phases that emerge in strongly correlated systems\cite{Wen:2002ez}.

Following this line of research, there has been a strong effort in describing and analysing a class of quantum states, known as stabilizer states, in the context of quantum error correction \cite{Gottesman:1997vn} and quantum computation \cite{Raussendorf:2001ys}. Surprisingly,  this very same class of states appears in a broad family of spin models that can describe topological quantum theories \cite{Levin:2005ve,Kitaev:2006ly,Bombin:2006zr}. As discovered by Kitaev, the understanding of these theories could provide new ideas about hardware for self-protected quantum memories and quantum computation \cite{Kitaev:2003qf,Freedman:2002bh}.

In the next sections, we will analyse a class of quantum states with a local tensor description. One of the most studied state with this local structure characterises the ground state of the Affleck-Kennedy-Lieb-Tasaki (\emph{AKLT}) model \cite{Affleck:1987cm,Affleck:1988uq}, a spin-1 chain with a Heisenberg-like Hamiltonian. The purpose of this work is to extend the \emph{AKLT} construction to higher dimensions and study the physical properties that emerge due to this local tensor structure from the quantum state.

To achieve these goals, several partial results will be presented in this work:

\begin{itemize}

\item An isotropic antiferromagnet spin-1 state is characterised just with symmetry arguments. This step can be seen as a generalisation of the \emph{AKLT} construction to two dimensions where we allow to have \emph{multipartite} bond states.

\item A parent spin-1 Hamiltonian is defined that has the multipartite valence bond state as a ground state. It is a four-body local and real Hamiltonian with explicit $SU(2)$ and parity invariance.

\item The correlations and structure of the multipartite valence bond ground state are analysed using a mapping to two dimensional classical statistical models and to one dimensional quantum systems. At this point, we will explore the connections of the antiferromagnet state with integrable models.

\item The physical relevance of the model will be motivated comparing it with the interaction and structures that appear in usual anti-ferromagnetic Mott-Hubbard insulators.

\end{itemize}

The paper is organised as follows: in the next section, we will describe the tensor structure of the ground state of the \emph{AKLT} model. This section will be useful not only to give an overview of the properties of valence bond states in one dimension but, at the same time, we will use it to introduce the notation for the rest of the paper. 

The third section contains the main body of this work. In this section, we will define the two dimensional multipartite valence bond state and obtain the parent Hamiltonian. We will then use a mapping to a two dimensional statistical model to characterise the properties of the ground state.

Finally, in the last section, we will give a physical motivation of this model. We will summarise some well known aspects and structures, like the hybridisation of the ionic orbitals by covalent mixing, that appears in anti-ferromagnetic insulators and that could be described by the valence bond structure of the state we are about to reveal.

\section{1D Valence Bond Ground States.}

Many of the features and structures that we will find in this work can be seen as a generalisation of the ground state of the \emph{AKLT} model \cite{Affleck:1987cm,Affleck:1988uq}, constructed from an antiferromagnet spin Hamiltonian with Heisenberg-like interaction between neighbour sites. The basic building block of this ground state is given by a structure, known as valence bond. Its inherent quantum nature has brought new perspectives to the field of low dimensional anti-ferromagnetic systems extending beyond the model itself. 

\begin{figure}[!ht]
\begin{center}
\resizebox{!}{2.0cm}{\includegraphics{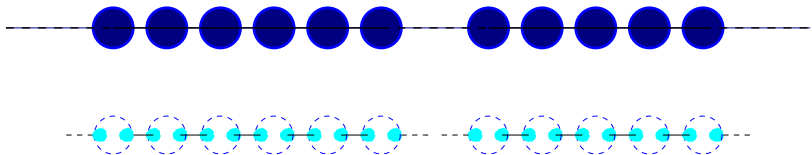}}
\caption[Valence bond ground state]{\label{aklt}Valence bond ground state. The first line with dark blue spots represents the physical state of the spin chain, while the second line represent the implementation of the state with two ancillae systems per site and a maximally entangled state between neighbour sites.}
\end{center}
\end{figure}

Moreover, several features make the quantum phase characterised by the \emph{AKLT} model, the Haldane phase \cite{Haldane:1983tg,Haldane:1983hc} important in its own right. First, the Heisenberg model is contained in this phase; this model is the subject of the Haldane conjecture which states that systems with half integer and integer spin have completely different behaviours. The former are gapless systems while the latter generate a mass gap that cannot be explained with perturbative methods and this is a pure quantum effect. Second, this phase is a magnetically disordered phase; it describes spin liquids with a hidden topological structure \cite{Girvin:1989fv} that cannot be represented with the usual two point correlators \cite{Nijs:1989ya,Verstraete:2004ps,Verstraete:2004qx,Popp:2005jb} (see Fig. \ref{aklt}).

Following the original works \cite{Affleck:1987cm,Affleck:1988uq}, the \emph{AKLT} model describes a translational invariant antiferromagnet spin-1 chain. Its construction is done in two steps:
\begin{enumerate}
\item Every local spin-1 system is constructed from the projection into the triplet subspace of two ancillary spin-$\frac{1}{2}$ subsystems. 
\item In the ancilla picture, two contiguous spin-$\frac{1}{2}$ of different sites are linked with a singlet state, i.e. a maximally entangled state with zero total spin angular momentum (valence bond).
\end{enumerate}
Denoting the state of a spin-$\frac{1}{2}$ subsystem by $|\alpha ) \in \mathbb{C}^2$, the second step in the construction of the \emph{VBS} fix the state between neighbouring ancilla spins to
\be
|0 ) =\sum_{\{\alpha,\beta\}=\{\uparrow,\downarrow\}} |\alpha) \epsilon_{\alpha \beta} |\beta ) = | \uparrow \downarrow ) - |\downarrow \uparrow ),
\ee
with $\epsilon_{\uparrow \downarrow}=-\epsilon_{\downarrow \uparrow}=1$ and $\epsilon_{\uparrow \uparrow}=-\epsilon_{\downarrow \downarrow}=0$. The projection of two spin-$\frac{1}{2}$ subsystems into the triplet subspace is imposed by
\be
|\psi_{\alpha  \beta} \rangle = \frac{1}{\sqrt{2}} \left( |\alpha) |\beta) + |\beta) |\alpha) \right) = 
\begin{cases} 
\sqrt{2} |+1 \ra & \alpha = \beta = \uparrow \\
|0\ra & \alpha \neq \beta \\
\sqrt{2} |-1 \ra & \alpha = \beta = \downarrow.
\end{cases}
\ee
In the first part of this equation appears the local state in a tensor notation, in the middle part, the state is described in the ancillae spin-$\frac{1}{2}$ space and in the last part of the equation, we find the same state written in the spin-1 language.

So that, if we plug in all the local tensor structures to define our state and using a basis in the spin-1 representation such that $S^{a}|b\rangle = i \epsilon^{abc} |c\rangle$, with $\{a,b,c\}=\{x,y,z\}$ and $\epsilon^{abc}$ being the Levi-Civita tensor, the ground state of the \emph{AKLT} model can be described locally as
\be
\left( |\psi \ra \cdot \epsilon \right)_{\alpha \gamma} =\sum_{s=\{x,y,z\}} \frac{1}{\sqrt{3}} \sigma^{s}_{\alpha \gamma} |s\rangle =\sum_{s=\{x,y,z\}} A_{\alpha \gamma}[s] |s\rangle  ,
\ee
where $\sigma^{s}$ are the usual Pauli matrices. This description corresponds to a matrix product state \cite{Kluemper:1993kl,Ostlund:1995kp}. The first point to notice is that the characterisation of the state with this particular set of tensors is one of many, due to the fact that we can always modify the tensor description with similarity transformations (gauge transformation). So, in what follows, we will refer to this particular choice of tensor to describe the triplet $SU(2)$ sector as \emph{symmetric gauge} description (for aesthetic reasons).

Several features emerge just because of the tensor structure of the state and it can be shown that these characteristics remain in the commensurate part of the Haldane phase.

\begin{itemize}

\item Infrared limit.- Although this model is not the infrared fixed point of the phase, it is exponentially close to it and contains the long range and topological properties of the Haldane phase. Following Verstraete and coauthors, a scale transformation can be performed in this class of states. At any point in the scale transformation, the local state is written by,
\be
|\Psi(\Lambda)\ra = \frac{\sqrt{1+3\Lambda}}{2} \sigma^0 |0\ra +i \frac{\sqrt{1-\Lambda}}{2}  \sum_{s=\{x,y,z\}} \sigma^s ~  |s \ra
\ee
where $|0\ra$ is the singlet state, $\{|x\ra, |y\ra, |z\ra \}$ form the triplet and the correlation length of the state is given by $\xi=\frac{-1}{\log{| \Lambda |}}$. Then, a step in the scale transformation is defined by
\be
|\Psi(\Lambda)\ra \to |\Psi(\tilde{\Lambda})\ra, ~~ \Lambda^2= \tilde{\Lambda}
\ee
As a consequence of this analysis, we can see how the entropy evolves for any given number of sites $L$ and correlation length $\xi=\frac{-1}{\log{|\Lambda|}}$,
\be
S(L,\Lambda)=3 \frac{\Lambda^{2^L}-1}{4} \log_2{ \frac{1-\Lambda^{2^L}}{4}} - \frac{1+3\Lambda^{2^L}}{4} \log_2{ \frac{1+3\Lambda^{2^L}}{4}}.
\ee 
Whenever $L\to \infty$ or $\xi \to 0$, the entropy goes to $S\to \log_2 4$ (see also \cite{Fan:2004ga}).

\item Compositions rules.- We have seen that the tensor structure of the state is due to the fact that we are representing the triplet sector of $SU(2)$ as the symmetric projection of two spin-$\frac{1}{2}$ subsystems. If we relax the projection and also allow the singlet to have a non zero amplitude of probability, the \emph{VBS} is just the decomposition of two spin-$\frac{1}{2}$ into invariant subspaces of $SU(2)$, i.e. $\frac{1}{2} \otimes \frac{1}{2} = 0 \oplus 1$. 

But reading the \emph{VBS} as a matrix product state, i.e. $|\alpha) = \sum_{s,\beta}  A_{\alpha \beta}[s] |s\rangle |\beta)$, we find that the composition of a spin-1 or spin-0 particle with a spin-$\frac{1}{2}$ is constrained, i.e. $\frac{1}{2} \otimes 0 = \frac{1}{2}$ and $\frac{1}{2} \otimes 1 = \frac{1}{2}$.

This constraint in the composition of the angular momentum describes an algebraic structure known as quantum groups (see for example \cite{Gomez:1996ly}). In this particular case the quantum group is labelled by $SU(2)_2$. The possible relation between the matrix product state construction and quantum groups was already suggested by Sierra and Martin-Delgado \cite{Sierra:1998fk} 

\item Non local properties.- Finally, we would like to comment about the relation between the hidden topological order of the Haldane phase \cite{Girvin:1989fv} and the potential application of this phase in quantum communication \cite{Briegel:1998fk}. Verstraete and coauthors \cite{Verstraete:2004ps,Verstraete:2004qx,Popp:2005jb} showed that the behaviour of the non-local string order parameter of den Nijs and Rommelse \cite{Nijs:1989ya} can be understood in terms of the special entanglement properties of the Haldane phase. Using the \emph{VBS} structure of the \emph{AKLT} model, they showed that it is possible to create a maximally entangled pair with local operations and local measurements at any distance. It can be shown that the \emph{VBS} structure does not change in the commensurate part of the Haldane phase and therefore any ground state in the phase can be used to obtain a maximally entangled state, at any distance, using local measurements. 

\end{itemize}


\section{ Two dimensional anti-ferromagnetic spin-1 system. The model}

In this section, we will characterise a quantum system in a two dimensional lattice following the valence bond construction of \emph{AKLT} \cite{Affleck:1987cm,Affleck:1988uq} in one dimension, but in a way complementary to their extension to two dimensions. The quantum system is characterised by a Hamiltonian that is composed of nearest neighbour plaquette interactions. Geometrically, it is homogeneous, translational and rotational invariant, i.e. invariant under all point group symmetries of the lattice where it is placed. The ground state of the four-body local Hamiltonian is a real singlet state of $SU(2)$, i.e. it is a non-chiral state respecting time-reversal and parity symmetries. The local Hilbert space, the physical degrees of freedom, are spin-1 particles that live on the links of a square lattice.

\begin{figure}[!ht]
\begin{center}
\resizebox{!}{3.0cm}{\includegraphics{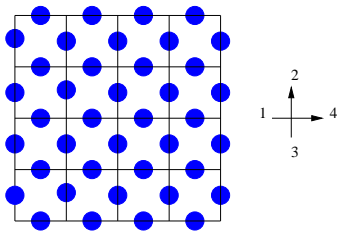}}
\caption{Lattice structure describing the model. At every bond, there is a spin-1 particle. The interactions take place around every vertex and involve the four nearest neighbour spins. If the physical degrees of freedom are located at the vertex of a square lattice, the system would be represented by a checkerboard lattice. The ground state of the system is characterised by a local tensor structure so a convention regarding the indexing of the tensors is needed. The picture shows the convention we are going to follow in this work, but it is totally arbitrary. The numbers are used to identify every spin involved in the interaction around the vertex.}
\end{center}
\end{figure}

The local interaction in this model appears around every vertex in the square lattice and involves up to four spin-1 particles. The structure of the Hamiltonian, that we are going to present, will be motivated at the end of this work where we will connect this model with the structure of the interactions that can appear in usual anti-ferromagnetic insulators (see for instance \cite{Imada:1998vn,Fazekas:1999kx} and references therein). In this work, we use a basis of the spin-1 representation such that $S^{\mu}|\sigma \rangle = i \epsilon^{\mu \sigma s} |s\rangle$, with $\{\mu, \sigma , s\}=\{x,y,z\}$ and $\epsilon^{\mu \sigma s}$ denoting the Levi-Civita tensor (the completely anti-symmetric tensor). The Hamiltonian for this system is described by:
\be
\begin{split}
H_{\text{vertex}} =& g_{1a} \left( \vec{S}_1 \cdot \vec{S}_2 + \vec{S}_2 \cdot \vec{S}_4 +\vec{S}_3 \cdot \vec{S}_4 +\vec{S}_1 \cdot \vec{S}_3 \right) + g_{1b} \left( \vec{S}_1 \cdot \vec{S}_4 +\vec{S}_2 \cdot \vec{S}_3 \right) \\
&+ g_{2a} \left( (\vec{S}_1 \cdot \vec{S}_2)^2 + (\vec{S}_2 \cdot \vec{S}_4)^2 +(\vec{S}_3 \cdot \vec{S}_4)^2 +(\vec{S}_1 \cdot \vec{S}_3)^2 \right) + g_{2b} \left( (\vec{S}_1 \cdot \vec{S}_4)^2 +(\vec{S}_2 \cdot \vec{S}_3)^2 \right) \\
&+g_3 \left( (\vec{S}_1 \cdot \vec{S}_4) (\vec{S}_2 \cdot \vec{S}_3) - (\vec{S}_1 \cdot \vec{S}_2) (\vec{S}_3 \cdot \vec{S}_4) - (\vec{S}_1 \cdot \vec{S}_3) (\vec{S}_2 \cdot \vec{S}_4)  \right) \\
&+g_{4a} \left( (\vec{S}_1 \cdot \vec{S}_2 + \vec{S}_3 \cdot \vec{S}_4) (\vec{S}_1 \cdot \vec{S}_3 +\vec{S}_2 \cdot \vec{S}_4) +  ( \vec{S}_1 \cdot \vec{S}_3 +\vec{S}_2 \cdot \vec{S}_4 )  (\vec{S}_1 \cdot \vec{S}_2 + \vec{S}_3 \cdot \vec{S}_4) \right) \\
&+g_{4b} \left(  (\vec{S}_1 \cdot \vec{S}_2 + \vec{S}_3 \cdot \vec{S}_4 +\vec{S}_1 \cdot \vec{S}_3 +\vec{S}_2 \cdot \vec{S}_4) ( \vec{S}_1 \cdot \vec{S}_4 +\vec{S}_2 \cdot \vec{S}_3 ) \right) \\
&+\Delta \sum_{\text{perm}\{1243\}} \left( (\vec{S}_1 \cdot \vec{S}_2)  (\vec{S}_2 \cdot \vec{S}_4)  (\vec{S}_4 \cdot \vec{S}_3)  (\vec{S}_3 \cdot \vec{S}_1) + (\vec{S}_1 \cdot \vec{S}_3)  (\vec{S}_3 \cdot \vec{S}_4)  (\vec{S}_4 \cdot \vec{S}_2)  (\vec{S}_2 \cdot \vec{S}_1)  \right)
\end{split}
\ee
The first term in the Hamiltonian corresponds to bilinear anti-ferromagnetic interactions between nearest neighbours; the second term describes an anti-ferromagnetic interaction between next nearest neighbours at every vertex. For the model we are going to study, its strength is about $g_{1b} \simeq 0.177 g_{1a}$. The third and fourth terms in the Hamiltonian represent similar interactions to the first and second terms but they are biquadratic operators. These kind of terms are allowed by symmetry considerations and by the local spin-1 representation of $SU(2)$ operators. Their amplitudes are about $g_{2a} \simeq 0.312 g_{1a}$ and $g_{2b} \simeq 0.034 g_{1a}$. The following terms are several multi-spin exchange interaction needed to stabilise the phase we are going to study. Their couplings are of the order of $g_{3} \simeq 0.056 g_{1a}$, $g_{4a} \simeq 0.165 g_{1a}$, $g_{4b} \simeq 0.120 g_{1a}$ and $\Delta \simeq 0.003 g_{1a}$.

The values of the coupling constants are a particular instance of a more general set of Hamiltonians and they just give an estimate. We will see in the following sections that the couplings in the Hamiltonian may, in fact, change quite drastically without changing the properties of the ground state. This fact gives a hint at the stability of the phase we are about to reveal. 

\subsection{The Ground State}

In what follows, we will see that the unique ground state of the Hamiltonian we have just presented corresponds to a valence bond state with a four-partite entangled state associated with the vertices of the lattice. The properties of the quantum system are inherited by the constraints imposed at the vertex and the dimensions of the local Hilbert space placed at every link. As a consequence, the parameters of the tensor structure that characterise the state will be completely defined by symmetry arguments. 

We are first looking for a uniform, translational invariant singlet state in the square lattice. Then, we restrict our attention to states of the form:
\begin{enumerate}
\item At every link, a local Hilbert space is defined by a projector of two ancillae subsystems in the spin-1/2 representation to the triplet representation of $SU(2)$.
\item At every vertex, four ancillae subsystems are entangled in a singlet of $SU(2)$. Following \cite{Wen:1989ys} two different states are allowed, depending on their chirality (left and right handed).
\end{enumerate}
\begin{figure}[!ht]
\begin{center}
\resizebox{!}{3.0cm}{\includegraphics{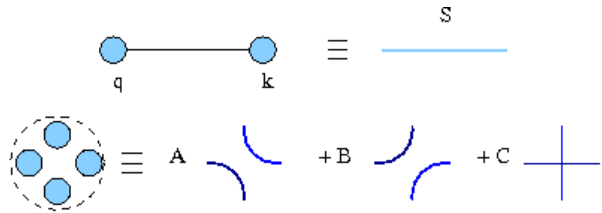}}
\caption{\label{ved} Representation of the physical Hilbert space defined at every link from the projector of two spin-1/2 ancillae subsystems and the entangled state at every vertex. In the analysed example, a singlet state is chosen at the vertex and it turns out to be a special case of the classical six vertex model.}
\end{center}
\end{figure}
Mathematically, both conditions are imposed using a set of tensors. First, at every link, the local quantum states are characterised by: 
\be
|\psi \ra = \sum_{s=\{x,y,z\}} a(s) ~ \sigma^s ~  |s \ra
\ee
with $a(s)$ some complex probability amplitude such that: $a(x)=a(y)=a(z)$ are the probability amplitudes of the triplet sector; and $\sigma^s$ correspond to the usual Pauli matrices.

At the vertices, four spin-$1/2$ meet to form a singlet state of $SU(2)$. Decomposing the Hilbert space spanned by these spins into invariant subspaces of $SU(2)$, i.e. $\frac{1}{2} \otimes \frac{1}{2} \otimes \frac{1}{2} \otimes \frac{1}{2} = 0_+ \oplus 0_- \oplus 1 \oplus 2$, we find that the singlet sector is not unique but it is two dimensional. This subspace was studied by Wen, Wilczek and Zee. They characterised it by an emergent feature of the four spin-$1/2$ states known as chirality. For example, the singlet sector with positive chirality is explicitly given by\footnote{We remind the reader that the physical degrees of freedom are given by spin-1 systems; the ancillary spin-$1/2$ system can be seen as a tool to represent the state. To emphasise the difference we will use curly and normal bra-kets , i.e. $|d) \in  \mathbb{C}^2$ and $|s\ra \in  \mathbb{C}^3$.}: 
\be
\begin{split}
|0_+) &= \frac{1}{\sqrt{6}} \left( |\uparrow \uparrow \downarrow \downarrow ) + | \downarrow \downarrow \uparrow \uparrow) + e^{i2\pi/3} [|\uparrow \downarrow \uparrow \downarrow ) + | \downarrow \uparrow \downarrow \uparrow) ]+ e^{-i2\pi/3} [| \downarrow \uparrow \uparrow \downarrow ) + | \uparrow \downarrow \downarrow \uparrow) ] \right) \\
&\equiv \sum_{\{a_1,b_2,c_3,d_4\} \in \{\uparrow,\downarrow\}} \Gamma_{a_1 c_3}^{b_2 d_4}[0_+] |a_1b_2c_3d_4)
\end{split}
\ee
The singlet state with negative chirality $|0_-)$ corresponds to the complex conjugate of $|0_+)$. The first point to notice is that these two states form a complete complex orthonormal basis for two dimensional singlet sector. Thus, any other state with zero total spin angular moment correspond to a superposition of these two states. For example, the state made out of two singlets pairs between the first spin and the second one and the third with the fourth spin can be rewritten as:
\be
|0_{12} 0_{34} ) =\frac{1}{\sqrt{2}} \sum_{\{a_1,b_2,c_3,d_4\} \in \{\uparrow,\downarrow\}} \left( \Gamma_{a_1 c_3}^{b_2 d_4}[0_+] - \Gamma_{a_1 c_3}^{b_2 d_4}[0_-] \right) |a_1b_2c_3d_4)
\ee
This equation is a particular instance of more general relations known as skein relations.

Due to the fact that we are looking for a real $SU(2)$ singlet, the entangled state that defines our system should be of the form:
\be
\Gamma_{a_1 c_3}^{b_2 d_4}[0]  =e^{i\phi} \Gamma_{a_1 c_3}^{b_2 d_4}[0_+] + e^{-i\phi} \Gamma_{a_1 c_3}^{b_2 d_4}[0_-],
\ee
a superposition of the two chiral sectors with equal weight.

If we rewrite the state in the vertex in the \emph{symmetric} gauge, we find that any real singlet state can be described by:
\be
\Gamma_{a_1 c_3}^{b_2 d_4}[0]  =  \cos{\phi} \, \sigma^0_{a_1 d_4}  \sigma^0_{c_3 b_2} 
+  \frac{\sin{\phi}}{\sqrt{3}} \left( \sigma^x_{a_1 d_4}  \sigma^x_{c_3 b_2} + \sigma^y_{a_1 d_4}  \sigma^y_{c_3 b_2} + \sigma^z_{a_1 d_4}  \sigma^z_{c_3 b_2} \right)
\ee
where $\phi$ is the relative angle between the positive and the negative chiral singlet sectors. 

It is straightforward to realise that the multipartite entangled state at every vertex corresponds to a special case of the classical six vertex model (see for instance \cite{Baxter:1989zr,Gomez:1996ly} and references therein), so that the state is decomposed in a classical structure ({\emph{scaffolding}}) defined at the vertex and the link where we place the triplet degrees of freedom.

So, in our problem of defining a $SU(2)$ singlet in a square lattice, we found that the vertex of the state is completely defined up to a phase. Then, we should impose another physically motivated restriction to fully characterised the multipartite entangled state. This constraint comes from the geometric properties of the lattice where the system is placed. To make explicit the rotational symmetry that we are going to impose in the state, we will not work with the chiral basis; instead we will decompose any possible singlet state of four spin-$1/2$ by the superposition of two singlets pairs between the first spin and the second one and the third with the fourth spin and another two singlets pairs between the first spin and the third one and the second with the fourth spin i.e. $\{|0_{12}) \otimes |0_{34}),|0_{13}) \otimes |0_{24}) \}$. This basis is a complete basis for the two dimensional singlet sector but it is not orthonormal. Nevertheless, writing the state in this basis makes explicit the properties of the state under rotation.

We are looking for a rotational invariant state, so the multipartite entangled state at every link should be an equal weighted superposition of the singlets $|0_{12}) \otimes |0_{34})$ and $|0_{13}) \otimes |0_{24})$, i.e.
\be
|0)  =e^{i\theta} |0_{12}) \otimes |0_{34}) + e^{-i\theta} |0_{13}) \otimes |0_{24})
\ee 

Summarising, the state that we are characterising is a real $SU(2)$ singlet that is rotational invariant. Only two solutions are possible with the underlying tensor structure we are using. Keeping the \emph{symmetric} gauge, the two solutions are given by:
\begin{enumerate}
\item A set of one dimensional structures along the rows and columns. They correspond to having unentangled \emph{AKLT} chains at every row and at every column. Mathematically the tensor that characterises the state at the vertex is given by
\be
\Gamma_{a_1 c_3}^{b_2 d_4}[0]^{(1)}  =  \sigma^0_{a_1 d_4}  \sigma^0_{c_3 b_2} 
\ee 

\item A multipartite entangled state of the four spin-$1/2$ ancillae subsystem such that the two states that come from the column combine in the triplet sector and then decay along the row. The tensor that represents this state is
\be
\Gamma_{a_1 c_3}^{b_2 d_4}[0]^{(2)}  =  \frac{1}{\sqrt{3}} \left( \sigma^x_{a_1 d_4}  \sigma^x_{c_3 b_2} + \sigma^y_{a_1 d_4}  \sigma^y_{c_3 b_2} + \sigma^z_{a_1 d_4}  \sigma^z_{c_3 b_2} \right)
\ee
\end{enumerate}

For the first solution, almost everything is known, due to the fact that it corresponds to unentangled one dimensional structures. We will not go on analysing the phase of this state but remind the reader that there are important one dimensional structures like stripes states in two dimensional anti-ferromagnetic insulators (see for example \cite{Kivelson:2003qf}). 

In the next sections, we will analyse and study the second possible solution.

\subsubsection*{Small remarks and generalisations.-} 

Once, we have characterised an isotropic singlet state in a square lattice, we can look for other models that generalise the $SU(2)$ symmetry. For example, in the hexagonal lattice, we can substitute every link with a projector of two three level systems that transform as $3\otimes \bar{3}$ into the octet, i.e $3\otimes \bar{3} \to 8 $. At every vertex, we will find three states $3\otimes 3 \otimes 3$ or $\bar{3}\otimes \bar{3} \otimes \bar{3}$. It happens that the $SU(3)$ singlet sector is unique and it is given by the Levi-Civita tensor of three indices. So that, the SU(3) singlet state in the hexagonal lattice is completely defined.

Looking at the geometric constraints in our model, the only requirement that we need in the structure of the lattice is that the connectivity of the lattice is of degree four. The square lattice is a particular instance that fulfils this requirement, another instance is the Kagome lattice. Nonetheless, in this case, the rotational symmetry on the state will come under $\pi/3$ rotations instead of the $\pi/2$ rotations of the square lattice.

\begin{figure}[!ht]
\begin{center}
\resizebox{!}{5.0cm}{\includegraphics{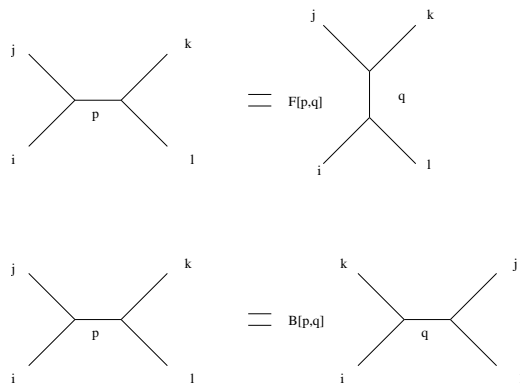}}
\caption{\label{fmatrix}Graphical representation of the unitary equivalent Hilbert space coming from the composition of the different subsystems in different order. The unitary that relates the Hilbert spaces is given by the classical Racah's symbols.}
\end{center}
\end{figure}

Finally, we would like to comment about the associativity properties and consistency condition of the tensor at every vertex. Up to now, we were just composing four spin-$1/2$ systems. If we carry on this composition in steps where we join or fuse two subsystems at a time the result can not depend in the way we choose the subsystems to fuse. This means that the Hilbert space spanned by the states where we fuse the first and the second with the third and fourth spin should be unitary equivalent to the Hilbert space spanned by the states where we fuse the first and the third with the second and the fourth spin. The unitary transformation that relates both equivalent spaces only depends on the total angular momentum of the states and in our case is given by the classical Racah's symbols (see for instance \cite{Galindo:1991ve} and references therein):
\be
F_{pq}\begin{bmatrix}\frac{1}{2} & \frac{1}{2} \\ \frac{1}{2} & \frac{1}{2} \end{bmatrix} = \begin{pmatrix} \frac{1}{2} & \frac{\sqrt{3}}{2} \\ \frac{\sqrt{3}}{2} & -\frac{1}{2} \end{pmatrix}
\ee
where $p$ and $q$ label the total spin angular momentum of the intermediate states in the composition of the four spin-$1/2$ systems (see Fig. \ref{fmatrix}).

\subsection{Local spin Hamiltonian}

The quantum system that we have just defined in the last section is characterised by the state: $\sum_{ \{ s \} } A[s] |s\ra= \frac{1}{\sqrt{3}} \sum_{s=\{x,y,z\}} \sigma^{(s)} |s\ra$ at every link and the tensor: $\Gamma[0]  =  \frac{1}{\sqrt{3}} \left( \sigma^x \otimes  \sigma^x + \sigma^y \otimes  \sigma^y + \sigma^z \otimes  \sigma^z \right)$ at every vertex. Around every vertex, this tensor structure spanned a sixteen dimensional subspace $\mathcal{K}$ from the eighty one dimensional Hilbert space of four spin-$1$. Any positive definite operator that involve the four spin-$1$ states around every vertex, and whose kernel is exactly the subspace $\mathcal{K}$, is a possible Hamiltonian, for which the valence bond state is the ground state. As we have seen, symmetry and geometric constraints fully specify the valence bond state. The freedom in choosing the Hamiltonian gives an idea of how stable this kind of valence bond phases are under any perturbation. In fact, the Hamiltonian need not respect the $SU(2)$ symmetry and still the ground state will be given by the valence bond construction.

Due to the interactions that usually appear in anti-ferromagnetic insulators, we find that the Hamiltonian given by
\be
\begin{split}
H_{\text{vertex}} =& g_{1a} \left( \vec{S}_1 \cdot \vec{S}_2 + \vec{S}_2 \cdot \vec{S}_4 +\vec{S}_3 \cdot \vec{S}_4 +\vec{S}_1 \cdot \vec{S}_3 \right) + g_{1b} \left( \vec{S}_1 \cdot \vec{S}_4 +\vec{S}_2 \cdot \vec{S}_3 \right) \\
&+ g_{2a} \left( (\vec{S}_1 \cdot \vec{S}_2)^2 + (\vec{S}_2 \cdot \vec{S}_4)^2 +(\vec{S}_3 \cdot \vec{S}_4)^2 +(\vec{S}_1 \cdot \vec{S}_3)^2 \right) + g_{2b} \left( (\vec{S}_1 \cdot \vec{S}_4)^2 +(\vec{S}_2 \cdot \vec{S}_3)^2 \right) \\
&+g_3 \left( (\vec{S}_1 \cdot \vec{S}_4) (\vec{S}_2 \cdot \vec{S}_3) - (\vec{S}_1 \cdot \vec{S}_2) (\vec{S}_3 \cdot \vec{S}_4) - (\vec{S}_1 \cdot \vec{S}_3) (\vec{S}_2 \cdot \vec{S}_4)  \right) \\
&+g_{4a} \left( (\vec{S}_1 \cdot \vec{S}_2 + \vec{S}_3 \cdot \vec{S}_4) (\vec{S}_1 \cdot \vec{S}_3 +\vec{S}_2 \cdot \vec{S}_4) +  ( \vec{S}_1 \cdot \vec{S}_3 +\vec{S}_2 \cdot \vec{S}_4 )  (\vec{S}_1 \cdot \vec{S}_2 + \vec{S}_3 \cdot \vec{S}_4) \right) \\
&+g_{4b} \left(  (\vec{S}_1 \cdot \vec{S}_2 + \vec{S}_3 \cdot \vec{S}_4 +\vec{S}_1 \cdot \vec{S}_3 +\vec{S}_2 \cdot \vec{S}_4) ( \vec{S}_1 \cdot \vec{S}_4 +\vec{S}_2 \cdot \vec{S}_3 ) \right) \\
&+\Delta \sum_{\text{perm}\{1243\}} \left( (\vec{S}_1 \cdot \vec{S}_2)  (\vec{S}_2 \cdot \vec{S}_4)  (\vec{S}_4 \cdot \vec{S}_3)  (\vec{S}_3 \cdot \vec{S}_1) + (\vec{S}_1 \cdot \vec{S}_3)  (\vec{S}_3 \cdot \vec{S}_4)  (\vec{S}_4 \cdot \vec{S}_2)  (\vec{S}_2 \cdot \vec{S}_1)  \right)
\end{split}
\ee
can be of some relevance in the studies of this kind of materials\cite{Imada:1998vn,Fazekas:1999kx}.

Using the parameter $\Delta >0$ as a perturbation, the operator $H_{\text{vertex}}(\Delta)$ is positive definite with a kernel (ground state) given by the valence bond state, if the other parameters are given by
\be
g_{1a}=1; ~~ g_{1b} = \frac{1+ 86 \Delta}{7}; ~~ g_{2a} = \frac{2}{7} \left(1+44 \Delta \right);~~ g_{2b}= \frac{446 \Delta -1}{7}; ~~ g_{3}= 20 \Delta; ~~ g_{4a}= \frac{3+167 \Delta}{21}; ~~ g_{4b}=\frac{3+734 \Delta}{42}  
\ee
Moreover, the gap to the first exited state of the local Hamiltonian in one vertex is given by: $\text{gap}= 360 \Delta$. So that, when the local gap is of order one, the parameters $g_i$ take the values displayed in the introduction.

Following the same steps as in the \emph{AKLT} construction \cite{Affleck:1987cm, Affleck:1988uq, Fannes:1990ur}, we can show that the valence bond state is, in fact, a unique ground state of $H_{\text{vertex}}$. Nevertheless, there is a sufficient condition to prove the uniqueness of the ground state \cite{Perez-Garcia:2007bh}. The condition states that if the mapping that goes from the Hilbert space spanned by the ancillae subsystem in the \emph{boundary} to the Hilbert space of the physical spin-$1$ degrees of freedom in the bulk is injective then the ground state is unique.

Once we have specified the state by the tensors $A[s] $ and $\Gamma[0]$ is straightforward to check the injectivity of the map and then, the uniqueness of the ground state.

\subsection{Ground state properties and correlations. Expectation values}

We now want to characterise the state we have just defined and to get its correlation values. The value of any expectation value is obtained via a mapping of the 2D quantum state to a 2D classical statistical model and from there to a 1d quantum mechanical problem using a transfer matrix defined from the 2D quantum state. The long distance properties of the state happens to be described by an integrable statistical model.

\subsubsection{2D Statistical Model.-}

In order to obtain any expectation value given a valence bond state, a vertex matrix $R$ is defined from the tensors at the links and at the vertices. So that, from these kind of valence bond models, two different classical structures appear in the vacuum-to-vacuum expectation value:
\begin{itemize}
\item At every bond, a transfer matrix is placed, defined by: $E=\sum_{s=\{x,y,z\}} |a(s)|^2 ~~ \left( \sigma^s \right)^* \otimes \sigma^s$
\item At every vertex, a vertex matrix is placed, defined by: $V=\Gamma^* \otimes \Gamma$
\end{itemize}
and the $R$ matrix is given by:
\be
R^{i_2,i_4}_{i_1,i_3} = \sum_{\{j_1,j_2,j_3,j_4\}} \sqrt{E_{i_1,j_1}} \sqrt{E_{i_3,j_3}} V^{j_2,j_4}_{j_1,j_3} \sqrt{E_{j_2,i_2}} \sqrt{E_{j_4,i_4}}
\ee

For instance, the vacuum to vacuum expectation value in the valence bond state correspond to the calculation of the classical partition function of the vertex model:
\be
\la \psi | \psi \ra = \sum_{\text{all configurations}} \prod_{\text{lattice}} R^{ij}_{lk}   = \mathcal{Z}_{2D}
\ee
where one sums over all possible configurations compatible with the weights $R$.
\begin{figure}[!ht]
\begin{center}
\resizebox{!}{4.0cm}{\includegraphics{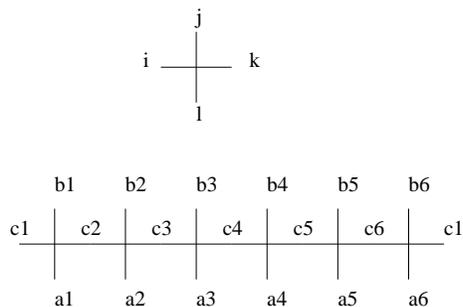}}
\caption{Vertex model in a classical model. The top part represents the vertex $R^{ij}_{lk} $, the Boltzmann weight of the vertex model. The bottom part represents the row-to-row transfer matrix for the vertex model.}
\end{center}
\end{figure}

So, given any valence bond state model described by local projectors in a two dimensional lattice, it is always possible to build a classical vertex model defined by a matrix $R$ that can be interpreted as Boltzmann weights of a classical model. Then, we have a mapping from a class of quantum states to classical models in a lattice with the same dimensions (see also \cite{Ardonne:2004fk,Verstraete:2006uq,Van-den-Nest:2007kx,Zhou:2007vn}).

Once, we have realised about this equivalence between 2D valence bond model and classical statistical models, we can use well-known tools from statistical mechanics to obtain information about the quantum state. For example, the sufficient conditions for the integrability of a classical model i.e. the star-triangle or Yang-Baxter equations \cite{Baxter:1989zr,Gomez:1996ly,Korepin:1997mz}, will become sufficient conditions to obtain information about the thermodynamic limit from the quantum valence bond state. Then, a possible strategy is to study the possible valence bond structures that give naturally integrable models. 

Instead of going on with the two dimensional classical analogy, we will use another well-known mapping, the equivalence between $D$-dimensional classical statistical models and $d=D-1$ dimensional quantum mechanical problems. This mapping is at the core of path integral \cite{Feynman:1965dq} and quantum Montecarlo methods \cite{Linden:1992cr}.

\subsubsection{1d Quantum System.-}

In quantum mechanical problems, {\emph{time}} evolution is carried out by the row-to-row transfer matrix $T$, with matrix elements $T_{ab}$ given by
\be
T_{ab} = R^{c_1b_1}_{a_1c_2}  R^{c_2b_2}_{a_2c_3} \cdots R^{c_nb_n}_{a_nb_1}.
\ee
So, if we understand this time evolution operator as the exponential of a quantum Hamiltonian $\hat{H}_{1d}=- \log{T}$ then the classical partition function $\mathcal{Z}_{2D}$ is just given by the trace of the density matrix of a Gibbs ensemble $\mathcal{Z}_{2D}=\tr{\left( e^{-N \hat{H}_{1d}} \right)}$.

If we are able to diagonalise the one dimensional quantum Hamiltonian $\hat{H}_{1d}$, we will be able to extract information about the 2D valence bond state. For example, consider the normalisation of the valence bond state,
\be
\la \psi | \psi \ra = \mathcal{Z}_{2D}=\tr{\left( e^{-N \hat{H}_{1d}} \right)} = \tr{ \left(\sum_{\mu \ge 1}  |\mu) (\lambda_{\mu})^N (\mu | \right) }
\ee
where $\{|\mu) ~ |~ \mu \in \mathbb{N} \}$ is a complete set of eigenvectors of the one dimensional quantum Hamiltonian and $\lambda_{\mu}$ are the eigenvalues of the transfer matrix $T$. Then, in the thermodynamic limit $N >> 1$, the value of the highest eigenvalue is fixed by normalisation to $\lambda_1 = 1$.

The other eigenvalues give information about the correlation in the system. For instance, the two point function in a 2D lattice is given by
\be
\begin{split}
\la O_{\vec{x}}  O_{\vec{0}} \ra & = ( 1| \tilde{O}_{\vec{x}} \left(\sum_{\mu}  |\mu) (\lambda_{\mu})^{|\vec{x}|-1} ( \mu | \right) \tilde{O}_{\vec{0}} |1) \\
&= \sum_{\mu} ( 1| \tilde{O}_{\vec{x}} |\mu ) ( \mu |\tilde{O}_{\vec{0}} |1) \left(\frac{\lambda_{\mu}}{|\lambda_{\mu}|}\right)^{|\vec{x}|-1} e^{-\frac{{|\vec{x}|-1}}{\xi_{\mu}}},
\end{split}
\ee
where $\xi_{\mu}=\frac{-1}{\log{|\lambda_{\mu}|}}$ defines the correlation length and $\tilde{O}_{\vec{x}}$ is the classical matrix similar to the classical transfer matrix $T$ but, in the $\vec{x}$-link, we substitute the matrix
\be
E=\sum_{s=\{x,y,z\}} |a(s)|^2 ~~ \left( \sigma^s \right)^* \otimes \sigma^s
\ee
by the matrix
\be
\sum_{\{s,\mu\}=\{x,y,z\}} a(s)^* a(\mu)  ~~ \left( \sigma^s \right)^* \otimes \sigma^{\mu} ~~  \langle s| O |\mu \rangle.
\ee

For the valence bond model that we are analysing, the local dimension of the classical indexes in the transfer matrix is four dimensional. This vector space can be understood as the direct sum of the singlet sector $|0)$ and the triplet sector $\{|x),|y),|z)\}$. To built the one dimensional quantum Hamiltonian from the normalisation of the 2D quantum state, we will use the spin operators in a four dimensional representation, such that,
\be
[J^k,J^p]=i\epsilon^{kpq} J^q, ~~~J^k |0) =0,~~~ J^k |p ) = i \epsilon^{kpq} |q),~~~ \{k,p,q\} \in \{x,y,z\}.
\ee
and the creation and annihilation operators,
\be
b^{\dagger}_k |q) =\delta_{q0} |k), ~~~b_k |p) =\delta_{kp} |0)~~~\{k,p\} \in \{x,y,z\}
\ee
With these definitions, the one dimensional quantum Hamiltonian can be recast into,
\be
\begin{split}
\hat{H}_{1d}=&\alpha_1 \vec{J}_n \cdot \vec{J}_{n+1} + \alpha_2 \left( \vec{J}_n \cdot \vec{J}_{n+1}  \right)^2  + \alpha_3 [ \left(\vec{J}_n \right)^2 +  \left( \vec{J}_{n+1} \right)^2 ] + \alpha_4 \left( \vec{J}_n \right)^2   \left( \vec{J}_{n+1} \right)^2 \\
+& \alpha_5 \sum_{k=\{x,y,z\}} \left( b^{\dagger}_{n,k} b_{n+1,k} + b_{n,k} b^{\dagger}_{n+1,k} \right) + \alpha_6  \sum_{k=\{x,y,z\}} \left( b^{\dagger}_{n,k} b^{\dagger}_{n+1,k} + b_{n,k} b_{n+1,k} \right)+ \alpha_7
\end{split}
\ee
with
\be
\begin{split}
\alpha_1=&\alpha_5 = \frac{\log{3}}{2};~~\alpha_6=\frac{3 \log{\left(\frac{497+136 \sqrt{13}}{81}\right)}}{4 \sqrt{13}};~~\alpha_7=  \frac{\log{\left(469176871410737 - 130126251314344 \sqrt{13}\right)}-20\log{3}}{4\sqrt{13}}+\log{3};\\
&\alpha_2=\frac{5 \log{\left(\frac{497+136 \sqrt{13}}{81}\right)}}{12 \sqrt{13}}-\frac{\log{3}}{6};~~\alpha_3=\frac{5 \log{\left(\frac{497+136 \sqrt{13}}{81}\right)}}{8 \sqrt{13}}+\frac{\log{3}}{4};~~\alpha_4=\frac{5 \log{\left(\frac{497-136 \sqrt{13}}{81}\right)}}{12 \sqrt{13}}+\frac{\log{3}}{6};
\end{split}
\ee

The first two terms in this one dimensional quantum Hamiltonian correspond to a bilinear-biquadratic anti-ferromagnetic interactions; the third, fourth and last term are diagonal terms that lift the onsite energy of the different states; the fifth and sixth are the hopping and pairing terms, respectively.

As we have seen, the spectrum of this one dimensional quantum Hamiltonian gives the correlation function of the 2D valence bond state. To solve it, we are going to use a numerical renormalization group method. For a related work where the renormalisation transformation is applied on the 2D lattice see \cite{Altman:2002la}.

But before going into the numerics, we would like to analyse from the structure of the 2D quantum state the extreme cases that we could obtain in our numerical study.

The 2D quantum state, as we have defined, has two structures: at every vertex, there is a tensor $\Gamma[0]  =  \frac{1}{\sqrt{3}} \left( \sigma^x \otimes  \sigma^x + \sigma^y \otimes  \sigma^y + \sigma^z \otimes  \sigma^z \right)$ that is equivalent to a particular instance of the classical six vertex model; at every link, there is quantum state $\sum_{ \{ s \} } A[s] |s\ra= \frac{1}{\sqrt{3}} \sum_{s=\{x,y,z\}} \sigma^{(s)} |s\ra$, that corresponds to a combination of states in the triplet sector. To analyses the extreme cases, we introduce a parameter $\Lambda$ dependence in the state at every link, such that
\be
|\psi(\Lambda) \ra =a(0) \sigma^0 |0\ra + \sum_{s=\{x,y,z\}} a(s) \sigma^s ~  |s \ra =\frac{\sqrt{1+3\Lambda}}{2} \sigma^0 |0\ra +i \frac{\sqrt{1-\Lambda}}{2}  \sum_{s=\{x,y,z\}} \sigma^s ~  |s \ra,
\ee
where the state $|0\ra$ represents the singlet state and $\{|x\ra,|y\ra,|z\ra \}$ the triplet sector. The scalars $a(0)$, $\{a(s)\}$ give the amplitude of probability in the singlet and triplet sectors, respectively.

When $\Lambda = -\frac{1}{3}$, we recover the spin-1 model we are dealing with. If $\Lambda=1$ the whole 2D valence bond state is decomposed in an uncorrelated classical tensor at every vertex and a product of singlets states at every link. And if $\Lambda=0$, the valence bond state is decomposed in a correlated classical tensor at every vertex and an equal superposition of the singlet and triplet sectors at every link. In this last case, although the classical and quantum structures are correlated and there is non-trivial entanglement, the whole quantum state has exactly zero correlation length. This feature is due to the fact that the transfer matrix that appears at every link when we calculate the vacuum to vacuum expectation value is given by,
\be
E=\sum_{\mu=\{0,x,y,z\}} |a(\mu)|^2 ~~ \left( \sigma^{\mu} \right)^* \otimes \sigma^{\mu}.
\ee
If we diagonalize this matrix we find one eigenvalue $e_0=1$ and three eigenvalues $e_x=e_y=e_z=\Lambda$. This means, when $\Lambda=0$, the transfer matrix is a one dimensional projector and any correlation length is zero (which does not mean that the state is trivial, since there are still non trivial quantum correlations, i.e. entanglement). Then, in principle, we could think to use the parameter $\Lambda$ to apply perturbation theory. The numerical results, that we will discuss in the next lines, show that when the $\Lambda$ insertion in the links is not a trivial one, the long range behaviour of the quantum system corresponds to a conformal invariant theory, i.e. an instance of an integrable model with an infinite number of conserved quantities and infinite correlation length. This result should be linked with a theorem proved by Hastings \cite{Hastings:2004gf} where he shows that if a state has an algebraic decay of the correlation function then the Hamiltonian for which this state is the ground state is gapless.

In the numerical analysis, we use a renormalization group method (CORE) \cite{Morningstar:1994nx,Morningstar:1996oq} that combines the decimation of the short distance or high energy degrees of freedom with a matrix product state ansatz \cite{Kluemper:1993kl,Ostlund:1995kp} for the ground state. The choice of the ansatz is due to the fact that the structure of the one dimensional Hamiltonian $\hat{H}_{1d}$ is equivalent to the interactions that appear in spin chains in the Haldane phase \cite{Haldane:1983tg,Haldane:1983hc}. There, we know, that the infrared limit of the phase is given by the \emph{AKLT} model, that is a special case of a matrix product state; and the phase transition is described in the continuum limit by a Wess-Zumino-Novikov-Witten (\emph{WZNW}) model on the $SU(2)$ group at the level $k=2$ \cite{Takhtajan:1982fk,Babujian:1983uq,Tsvelik:1990kx}.

\begin{figure}[!ht]
\begin{center}
\resizebox{!}{3.0cm}{\includegraphics{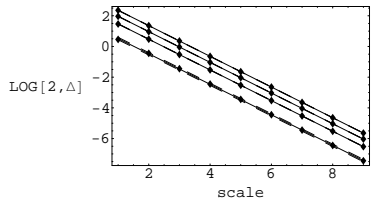}}
\caption{\label{gap} Plots of the logarithm in base two of the gap from the one dimensional quantum Hamiltonian as short distance degrees of freedom are integrated. By scale, we mean the number of steps in the renormalization, where we take groups of two contiguous sites every step. The points correspond to the numerical data, from top to bottom, they belong to $\Lambda=\{1/100,1/30,1/10,1/3\}$, respectively. The solid line corresponds to the fitted curve and the dashed line to a $95\%$ confidence interval for the predicted responses given a linear fit. The slope of the curve corresponds to a critical exponent  $\theta \simeq -0.99(4)$.}
\end{center}
\end{figure}

\begin{figure}[!ht]
\begin{center}
\resizebox{!}{3.0cm}{\includegraphics{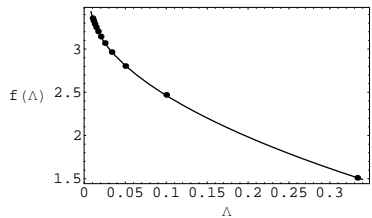}}
\caption{\label{constant} Plot of the function $f(\Lambda)$. The points correspond to the numerical data in the interval $\Lambda \in [1/100, 1/3]$, the line is just to guide the eye following the function $f(\Lambda) \simeq \frac{0.1}{\Lambda^{0.4}} + 1.1 \Lambda^{0.3} |\log_2{\Lambda}|^{1.1}$. For more details, look in the main text.}
\end{center}
\end{figure}

Due to theoretical analysis and scaling arguments, we know that, in this conformal model, the minimum gap  behaves as $\Delta(\Lambda) \simeq N^{\theta} 2^{f(\Lambda)}$, with $\theta=-1$, as we modify the infrared cut-off $N=2^{\text{scale}}$. Fig.\ref{gap} shows the plots for the behaviour of the minimum gap. The slopes in the fitted curves show the scaling of $\Delta(\Lambda)$ as the infrared cut-off is modified with $\theta \simeq -0.99(4)$. The plots correspond to values of $\Lambda=\{1/100,1/30,1/10,1/3\}$.

In the second plot, Fig.\ref{constant}, we check how change the starting point in the minimum gap as we change the value of $\Lambda \in [1/100, 1/3]$. The fit curve is just to guide the eye but follows a functional form $f(\Lambda) \simeq \frac{\alpha_1}{\Lambda^{\beta_1}} + \alpha_2 \Lambda^{\beta_2} |\log_2{\Lambda}|^{\beta_3}$ for some constant values $\{ \alpha_i,\beta_j \}$ given in the footnote of Fig.\ref{constant}.

Before closing this section, we would like to point out the connections between some effective (\emph{gauge}) theories and the features of the model we have just revealed. The structure of this anti-ferromagnet is given by a vertex tensor $\Gamma[0]$ and a link state $|\psi(\Lambda) \ra =\frac{\sqrt{1+3\Lambda}}{2} \sigma^0 |0\ra +i \frac{\sqrt{1-\Lambda}}{2}  \sum_{s=\{x,y,z\}} \sigma^s ~  |s \ra$. When the insertion of the parameter $\Lambda=0$, all the correlations in the system are zero, i.e. the Hilbert space of the theory at the boundary is one dimensional, the local Hamiltonian and dynamics can be expressed by a set of commuting projectors, i.e. local constraints, and nevertheless it can be said that the theory is by no means trivial. As soon as the insertion of the parameter $\Lambda$ is different from zero the theory that appears at the boundary of the two dimensional system is given by a very specific integrable model.

In 1989, Witten \cite{Witten:1989vn, Witten:1989ys, Witten:1990zr} noted that the boundary behaviour of a class of quantum field theories known as Chern-Simons theories in $2+1$ dimensions are closely tied to a class of conformal field theories called Wess-Zumino-Novikov-Witten (\emph{WZNW}) models in $1+1$ \cite{Belavin:1984vg,Moore:1989ly}. Among many other things, Witten showed that the Hilbert space at the boundary in these $2+1$ theories is one dimensional if there is no inclusion of Wilson lines. As soon as this inclusion is not zero, it starts to appears the whole Hilbert space structure of the conformal theories (see also \cite{Fradkin:1998ve,:1990bh} and reference therein). 
 
If the $\Lambda$ insertions in our microscopic model are understood as inclusions of Wilson lines, the system that we have just defined and characterised can be understood as a microscopic model of a Chern-Simons theory in the bulk with a $SU(2)_2$ \emph{WZNW} model living on the edge.

It might be interesting to go deeper into the analogies that appear between the microscopic model, we have studied, and Chern-Simons theories. For instance, features like the degeneracy of the Hilbert space depending on the topology where is defined should, in principle, be understood from the microscopic anti-ferromagnet.

\section{Anti-ferromagnetic Mott-Hubbard insulators}

The purpose of this section is not to make any claims about the understanding of the physics of anti-ferromagnetic Mott-Hubbard insulators. It merely shows that the ground state structure of these systems could be described by a valence bond states in two dimension. A complete comparison with experiments is needed (see for example \cite{Christensen:2007it} and references therein).

Most anti-ferromagnetic Mott-Hubbard insulators are transition metal compounds, in which the \emph{d}-orbital cations are separated by large anions \cite{Imada:1998vn,Fazekas:1999kx}. An example of these anti-ferromagnets is given by ceramic cuprate systems, basic building blocks of high-temperature superconductors. These cuprates are build of layers of $CuO_2$ where the copper cations are placed at the vertices of a square lattice separated by oxygen ions. Electrons in the last shell of the copper has a partially filled $3d^9$ configuration, while the oxygen sites are completely filled. Direct hopping between \emph{d}-orbitals seems rather unlikely.

Anderson \cite{Anderson:1950dq} showed that the concept of kinetic exchange is enhanced by mixing between cation and anion orbitals, explaining the large ordering temperatures of these materials. The anion-mediated exchange described in what follows is known as super-exchange. To understand how it works, we will consider a simple model that contains the essential features to explain the physics in this system. 

The model is made out of two cations with one \emph{d}-electron each and the intervening anion is an $O^{2-}$ anion. In the ionic picture, i.e. when there is no overlap between the orbital wave-functions of the ions, the oxygen has a filled \emph{p}-shell. As soon as the \emph{d} and \emph{p} orbitals overlaps, covalent mixing allows the \emph{p}-electrons to partially reoccupy the cations. Following \cite{Sawatzky:1976cr} the problem is treated from the point of view of the \emph{p}-electrons, for the two cases when the \emph{d}-spins are either in the triplet (parallel) or in the singlet (anti-parallel) configuration.

The on-site energies of the \emph{p} and \emph{d} orbitals are given by $E_p=\la \sigma_p | \mathcal{H} | \sigma_p \ra$ and $E_d=\la \sigma_d | \mathcal{H} | \sigma_d \ra$, with $\sigma=\{\uparrow, \downarrow \}$. The atomic levels are understood to arise from the solution of the many-electron problem at the ionic level. We assume that $E_p < E_d$. The covalent mixing amplitude of the \emph{p}-\emph{d} orbitals is given from perturbation theory by $\lambda \simeq \frac{\la \sigma_p | \mathcal{H} | \sigma_d \ra}{E_p - E_d}$.

\begin{figure}[!ht]
\begin{center}
\resizebox{!}{5.0cm}{\includegraphics{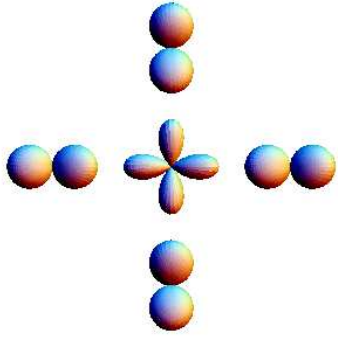}}
\caption{Lattice structure of usual antiferromagnet insulator. The anion is place at the link of the lattice with a \emph{p}-orbital configuration. The cation is place at the vertex of the lattice in a \emph{d}-orbital configuration. Super-exchange between two cations takes place via an intermediate anion. In the ionic picture, the anion has the outermost shell filled. Covalent mixing allows the anion \emph{p} electrons to partially reoccupy the cation \emph{d} orbitals. The mixing depends on the spin configuration.}
\end{center}
\end{figure}

If both \emph{d}-spins are parallel, e.g. $|\uparrow \uparrow \ra$, the $\downarrow$-spin \emph{p}-electron can extend to the cations on its left and right side, while the $\uparrow$-spin \emph{p}-electron must stay at the anion due to the Pauli exclusion principle,
\be
\begin{split}
| \downarrow_p \ra ~ \, ~ &\underrightarrow{\text{covalent-mixing}} ~ \, ~ \frac{| \downarrow_p \ra + \lambda | \downarrow_{d_L} \ra + \lambda | \downarrow_{d_R} \ra}{\sqrt{1+2\lambda^2}}  \\
| \uparrow_p \ra ~ \, ~ &\underrightarrow{\text{Pauli principle}} ~ \, ~ | \uparrow_p \ra.
\end{split}
\ee 
The energy of the two \emph{p}-electrons is given by
\be
\begin{split}
E_{\uparrow \uparrow} & \simeq \frac{1}{1+2\lambda^2} [ \left( \la \downarrow_p | + \lambda \la \downarrow_{d_L} | + \lambda \la \downarrow_{d_R} | \right)  \la \uparrow_p | \mathcal{H} | \uparrow_p \ra \left( | \downarrow_p \ra + \lambda | \downarrow_{d_L} \ra + \lambda | \downarrow_{d_R} \ra \right)  ] \\
&= 2 E_p + \frac{2 \lambda^2}{1+ 2 \lambda^2} (E_p - E_d).
\end{split}
\ee
If the \emph{d}-electron spin configurations is $|\downarrow \uparrow \ra$, the $\downarrow$-spin \emph{p}-electron can extend to the right cation and the $\uparrow$-spin \emph{p}-electron to the left one,
\be
\begin{split}
| \downarrow_p \ra ~ \, ~ &\underrightarrow{\text{covalent-mixing}} ~ \, ~ \frac{| \downarrow_p \ra  + \lambda | \downarrow_{d_R} \ra}{\sqrt{1+\lambda^2}}  \\
| \uparrow_p \ra ~ \, ~ &\underrightarrow{\text{covalent-mixing}} ~ \, ~ \frac{| \uparrow_p \ra + \lambda | \uparrow_{d_L} \ra }{\sqrt{1+\lambda^2}}.
\end{split}
\ee 
The total \emph{p}-electron energy contribution is estimated as
\be
\begin{split}
E_{\downarrow \uparrow} &\simeq \frac{1}{\left( 1+\lambda^2 \right)^2} [ \left( \la \downarrow_p |  + \lambda \la \downarrow_{d_R} | \right)  \left( \la \uparrow_p |  + \lambda \la \uparrow_{d_L} | \right) \mathcal{H} \left( | \uparrow_p \ra  + \lambda | \uparrow_{d_L} \ra \right) \left( | \downarrow_p \ra  + \lambda | \downarrow_{d_R} \ra \right)  ] \\
&=2 E_p + \frac{2 \lambda^2}{1+ \lambda^2} (E_p - E_d).
\end{split}
\ee
The super-exchange splitting is approximately
\be
E_{\uparrow \uparrow}-E_{\downarrow \uparrow} \simeq \frac{2 \lambda^4 (E_d - E_p)}{(1+2\lambda^2)(1+\lambda^2)}.
\ee
The parallel configuration costs more energy, thus it is more probable to find the system in the singlet configuration than the triplet one at every link. As soon as the orbital wave functions of the different ions start to overlap, a splitting between the singlet and triplet sector appears.

In the valence bond state, we have found that the structure of the state at the vertex of the 2D lattice is completely defined by global symmetries. The vertex tensor
\be
\Gamma[\phi]=\cos{\phi} \, \sigma^0  \sigma^0 +  \frac{\sin{\phi}}{\sqrt{3}} \sum_{s=\{x,y,z\}}  \sigma^s  \sigma^s
\ee
defines locally a real singlet state of $SU(2)$. The structure of the state at the link depends on a parameter $\Lambda$ that describes the splitting in the amplitude of probability of finding the system in a triplet or singlet configuration; when $\Lambda=0$ both sectors have the same amplitude of probability and any correlation in the system is zero. As soon as $\Lambda \neq 0$ there is a singlet-triplet splitting and the 2D anti-ferromagnet starts to develop correlations.










\section{Acknowledgements}

The authors thank the fruitful discussions at the QIG group in Innsbruck, where this project started and was finished. Also, the authors acknowledge comments and suggestions that appeared in several conversations with G. Brennen, A. Carollo, M.A. Martin-Delgado, G. Ortiz, U. Schollwoeck, F. Verstraete and M. Wolf. The authors thank the careful reading of the manuscript and valuable remarks by W. D\"ur, B. Kraus and M. Van den Nest. E.R. would like to thank ICTP, Trieste for hospitality during the workshops on \emph{Quantum Magnetism} and \emph{Cold Atomic Physics}, where part of this work was done and where we learnt about very interesting works \cite{Lauchli:2006ys,Schroeter:2007zr} about spin-1 quantum systems. These works discuss yet other phases of this system as quadrupolar or chiral phases.

Specially, E.R. thanks the support and encouraging conversations with M. Almeida, S. Fdez.-Vidal, K. Hammerer, S. Montangero, R. Rodriquez and L. Viola.

This work was supported by the Austrian Science Foundation (FWF), and the European Union (OLAQUI,SCALA,QICS).

\bibliography{link}

\begin{thebibliography}{10}

\bibitem{Bardeen:1957mv}
J.~Bardeen, L.~N. Cooper, and J.~R. Schrieffer.
\newblock Theory of superconductivity.
\newblock {\em Phys. Rev.}, 108:1175--1204, 1957.

\bibitem{Laughlin:1983fy}
R.~B. Laughlin.
\newblock Anomalous quantum hall effect: An incompressible quantum fluid with
  fractionally charged excitations.
\newblock {\em Phys. Rev. Lett.}, 50:1395, 1983.

\bibitem{Fradkin:1991ho}
E.~Fradkin.
\newblock {\em Field Theories of Condensed Matter systems}.
\newblock Addison-Wesley, 1991.

\bibitem{Auerbach:1994yp}
A.~Auerbach.
\newblock {\em Interacting electrons and Quantum Magnetism}.
\newblock Springer, 1994.

\bibitem{Essler:2005ly}
F.~H.~L. Essler, H.~Frahm, F.~G{\"o}hmann, A.~Kl{\"u}mper, and V.~E. Korepin.
\newblock {\em The One-Dimensional Hubbard Model}.
\newblock Cambridge University Press, 2005.

\bibitem{Rokhsar:1988uq}
D.S. Rokhsar and S.A. Kivelson.
\newblock Superconductivity and the quantum hard-core dimer gas.
\newblock {\em Phys. Rev. Lett.}, 61:2376, 1988.

\bibitem{Moessner:2001kx}
R.~Moessner and S.L. Sondhi.
\newblock Resonating valence bond phase in the triangular lattice quantum dimer
  model.
\newblock {\em Phys. Rev. Lett.}, 86:1881, 2001.

\bibitem{Misguich:2005fk}
G.~Misguich and C.~Lhuillier.
\newblock {\em Frustrated spin systems}.
\newblock World Scientific, 2005.

\bibitem{Anderson:1987oq}
P.W. Anderson.
\newblock The resonating valence bond state in la$_2$cuo$_4$ and
  superconductivity.
\newblock {\em Science}, 235:1196, 1987.

\bibitem{Zhang:1988nx}
F.~C. Zhang and T.M. Rice.
\newblock Effective hamiltonian for the superconducting cu oxides.
\newblock {\em Phys. Rev.}, B37:3759, 1988.

\bibitem{Ramirez:1997fk}
A.P. Ramirez.
\newblock Colossal magnetoresistance.
\newblock {\em J. Phys.: Condens. Matter}, 9:8171, 1997.

\bibitem{Bennett:2000at}
C.H. Bennett and D.P. DiVincenzo.
\newblock Quantum information and computation.
\newblock {\em Nature}, 404:247, 2000.

\bibitem{Nielsen:2000ne}
M.~A. Nielsen and I.~L. Chuang.
\newblock {\em Quantum computation and quantum information}.
\newblock Cambridge Univ. Press, 2000.

\bibitem{:2001fz}
Entanglement: theory and experiment (special issue).
\newblock {\em Quant. Inf. and Comp.}, 1, 2001.

\bibitem{Esteve:2003uo}
D.~Est{\`e}ve, J.M. Raimond, and J.~Dalibard, editors.
\newblock {\em Quantum entanglement and information processing}.
\newblock Les Houches, Session LXXIX, 2003.

\bibitem{Amico:2007fk}
L.~Amico, R.~Fazio, A.~Osterloh, and V.~Vedral.
\newblock Entanglement in many-body systems.
\newblock {\em Rev. Mod. Phys.}, pages quant--ph/0703044, 2007.

\bibitem{Preskill:1999he}
J.~Preskill.
\newblock Quantum information and physics: Some future directions.
\newblock {\em J. Mod. Opt.}, 47:127--137, 2000.

\bibitem{Wen:2002ez}
X.G. Wen.
\newblock Quantum order: a quantum entanglement of many particles.
\newblock {\em Phys. Lett.}, A300:175, 2002.

\bibitem{Gottesman:1997vn}
D.~Gottesman.
\newblock {\em Stabilizer Codes and Quantum Error Correction}.
\newblock PhD thesis, Caltech, 1997.

\bibitem{Raussendorf:2001ys}
R.~Raussendorf and H.~J. Briegel.
\newblock A one-way quantum computer.
\newblock {\em Phys. Rev. Lett.}, 86:5188, 2001.

\bibitem{Levin:2005ve}
M.A. Levin and X.G. Wen.
\newblock String-net condensation: A physical mechanism for topological phases.
\newblock {\em Phys. Rev.}, B71:045110, 2005.

\bibitem{Kitaev:2006ly}
A.~Kitaev.
\newblock Anyons in an exactly solved model and beyond.
\newblock {\em Annals of Phys.}, 321:2, 2006.

\bibitem{Bombin:2006zr}
H.~Bombin and M.~A. Martin-Delgado.
\newblock Topological quantum distillation.
\newblock {\em Phys. Rev. Lett.}, 97:180501, 2006.

\bibitem{Kitaev:2003qf}
A.~Kitaev.
\newblock Fault-tolerant quantum computation by anyons.
\newblock {\em Annals of Phys.}, 303:2, 2003.

\bibitem{Freedman:2002bh}
M.H. Freedman, A.~Kitaev, M.J. Larsen, and Z.~Wang.
\newblock Topological quantum computation.
\newblock {\em Bull. American Math. Soc.}, 40:31, 2002.

\bibitem{Affleck:1987cm}
I.~Affleck, T.~Kennedy, E.~H. Lieb, and H.~Tasaki.
\newblock Rigorous results on valence-bond ground states in antiferromagnets.
\newblock {\em Phys. Rev. Lett.}, 59:799, 1987.

\bibitem{Affleck:1988uq}
I.~Affleck, T.~Kennedy, E.~H. Lieb, and H.~Tasaki.
\newblock Valence bond ground states in isotropic quantum antiferromagnets.
\newblock {\em Commun. Math. Phys.}, page 477, 1988.

\bibitem{Haldane:1983tg}
F.D.M. Haldane.
\newblock Continuum dynamics of the 1-d heisenberg antiferromagnet:
  Identification with the o(3) nonlinear sigma model.
\newblock {\em Phys. Lett.}, A93:464, 1983.

\bibitem{Haldane:1983hc}
F.D.M. Haldane.
\newblock Nonlinear field theory of large-spin heisenberg antiferromagnets:
  Semiclassically quantized solitons of the one-dimensional easy-axis neel
  state.
\newblock {\em Phys. Rev. Lett.}, 50:1153, 1983.

\bibitem{Girvin:1989fv}
S.M. Girvin and D.P. Arovas.
\newblock Hidden topological order in integer quantum spin chains.
\newblock {\em Physica Scripta}, T(27):156, 1989.

\bibitem{Nijs:1989ya}
M.~den Nijs and K.~Rommelse.
\newblock Preroughening transitions in crystal surfaces and valence-bond phases
  in quantum spin chains.
\newblock {\em Phys. Rev}, B40:4709, 1989.

\bibitem{Verstraete:2004ps}
F.~Verstraete, M.~Popp, and J.~I. Cirac.
\newblock Entanglement versus correlations in spin systems.
\newblock {\em Phys. Rev. Lett.}, 92:027901, 2004.

\bibitem{Verstraete:2004qx}
F.~Verstraete, M.~A. Martin-Delgado, and J.~I. Cirac.
\newblock Diverging entanglement length in gapped quantum spin systems.
\newblock {\em Phys. Rev. Lett.}, 92:087201, 2004.

\bibitem{Popp:2005jb}
M.~Popp, F.~Verstraete, M.~A. Martin-Delgado, and J.~I. Cirac.
\newblock Localizable entanglement.
\newblock {\em Phys. Rev.}, A71:042306, 2005.

\bibitem{Kluemper:1993kl}
A.~Kluemper, A.~Schadschneider, and J.~Zittartz.
\newblock Matrix-product-groundstates for one-dimensional spin-1 quantum
  antiferromagnets.
\newblock {\em Europhys. Lett.}, 24:293, 1993.

\bibitem{Ostlund:1995kp}
S.~Ostlund and S.~Rommer.
\newblock Thermodynamic limit of density matrix renormalization.
\newblock {\em Phys. Rev. Lett.}, 75:3537, 1995.

\bibitem{Fan:2004ga}
H.~Fan, V.~Korepin, and V.~Roychowdhury.
\newblock Entanglement in a valence-bond-solid state.
\newblock {\em Phys. Rev. Lett.}, 93(22):227203, 2004.

\bibitem{Gomez:1996ly}
C.~Gomez, M.~Ruiz-Altaba, and G.~Sierra.
\newblock {\em Quantum Groups in Two-Dimensional Physics}.
\newblock Cambridge Univ. Press, 1996.

\bibitem{Sierra:1998fk}
G.~Sierra and M.~A. Martin-Delgado.
\newblock The density matrix renormalization group, quantum groups and
  conformal field theory.
\newblock {\em arXiv:cond-mat/9811170v3}, 1998.

\bibitem{Briegel:1998fk}
H.~J. Briegel, W.~Dur, J.~I. Cirac, and P.~Zoller.
\newblock Quantum repeaters: The role of imperfect local operations in quantum
  communication.
\newblock {\em Phys. Rev. Lett.}, 81:5932, 1998.

\bibitem{Imada:1998vn}
M.~Imada, A.~Fujimori, and Y.~Tokura.
\newblock Metal-insulator transitions.
\newblock {\em Rev. Mod. Phys.}, 70:1039, 1998.

\bibitem{Fazekas:1999kx}
P.~Fazekas.
\newblock {\em Lecture Notes on Electron Correlation and Magnetism}.
\newblock World Scientific, 1999.

\bibitem{Wen:1989ys}
X.G. Wen, F.~Wilczek, and A.~Zee.
\newblock Chiral spin states and superconductivity.
\newblock {\em Phys. Rev.}, B39:11413, 1989.

\bibitem{Baxter:1989zr}
B.J. Baxter.
\newblock {\em Exactly Solved Models in Statistical Mechanics}.
\newblock Academic Press, 1989.

\bibitem{Kivelson:2003qf}
S.A. Kivelson, I.P. Bindloss, E.~Fradkin, V.~Oganesyan, J.M. Tranquada,
  A.~Kapitulnik, and C.~Howald.
\newblock How to detect fluctuating stripes in the high-temperature
  superconductors.
\newblock {\em Rev. Mod. Phys.}, 75:1201, 2003.

\bibitem{Galindo:1991ve}
A.~Galindo and P.~Pascual.
\newblock {\em Quantum Mechanics}.
\newblock Springer, 1991.

\bibitem{Fannes:1990ur}
M.~Fannes, B.~Nachtergaele, and R.~F. Werner.
\newblock Finitely correlated states on quantum spin chains.
\newblock {\em Commun. Math. Phys.}, 144:443--490, 1992.

\bibitem{Perez-Garcia:2007bh}
D.~Perez-Garcia, F.~Verstraete, J.~I. Cirac, and M.~M. Wolf.
\newblock Peps as unique ground states of local hamiltonians.
\newblock {\em arXiv:0707.2260v1}, 2007.

\bibitem{Ardonne:2004fk}
E.~Ardonne, P.~Fendley, and E.~Fradkin.
\newblock Topological order and conformal quantum critical points.
\newblock {\em Annals of Phys.}, 310:493, 2004.

\bibitem{Verstraete:2006uq}
F.~Verstraete, M.~M. Wolf, D.~Perez-Garcia, and J.~I. Cirac.
\newblock Criticality, the area law, and the computational power of projected
  entangled pair states.
\newblock {\em Phys. Rev. Lett.}, 96:220601, 2006.

\bibitem{Van-den-Nest:2007kx}
M.~Van~den Nest, W.~Dur, and H.~J. Briegel.
\newblock Classical spin models and the quantum-stabilizer formalism.
\newblock {\em Phys. Rev. Lett.}, 98:117207, 2007.

\bibitem{Zhou:2007vn}
H.Q. Zhou, R.~Orus, and G.~Vidal.
\newblock Ground state fidelity from tensor network representations.
\newblock {\em arXiv:0709.4596v1}, 2007.

\bibitem{Korepin:1997mz}
V.~E. Korepin, N.~M. Bogoliubov, and A.~G. Izergin.
\newblock {\em Quantum Inverse Scattering Method and Correlation Functions}.
\newblock Cambridge University Press, 1997.

\bibitem{Feynman:1965dq}
R.P. Feynman and A.R. Hibbs.
\newblock {\em Quantum Mechanics and Path Integrals}.
\newblock McGraw-Hill Co., 1965.

\bibitem{Linden:1992cr}
W.~von~der Linden.
\newblock A quantum monte carlo approach to many-body physics.
\newblock {\em Phys. Rept.}, 220:53, 1992.

\bibitem{Altman:2002la}
E.~Altman and A.~Auerbach.
\newblock Plaquette boson-fermion model of cuprates.
\newblock {\em Phys. Rev.}, B65:104508, 2002.

\bibitem{Hastings:2004gf}
M.~B. Hastings.
\newblock Lieb-schultz-mattis in higher dimensions.
\newblock {\em Phys. Rev.}, B69:104431, 2004.

\bibitem{Morningstar:1994nx}
C.J. Morningstar and M.~Weinstein.
\newblock Contractor renormalization group method: A new computational
  technique for lattice systems.
\newblock {\em Phys. Rev. Lett.}, 73:1873, 1994.

\bibitem{Morningstar:1996oq}
C.J. Morningstar and M.~Weinstein.
\newblock Contractor renormalization group technology and exact hamiltonian
  real-space renormalization group transformations.
\newblock {\em Phys. Rev.}, D54:4131, 1996.

\bibitem{Takhtajan:1982fk}
L.A. Takhtajan.
\newblock The picture of low-lying excitations in the isotropic heisenberg
  chain of arbitrary spins.
\newblock {\em Phys. Lett.}, A87:479, 1982.

\bibitem{Babujian:1983uq}
H.M. Babujian.
\newblock Exact solution of the isotropic heisenberg chain with arbitrary
  spins: Thermodynamics of the model.
\newblock {\em Nucl. Phys.}, B215:317, 1983.

\bibitem{Tsvelik:1990kx}
A.M. Tsvelik.
\newblock Field-theory treatment of the heisenberg spin-1 chain.
\newblock {\em Phys. Rev.}, B42:10499, 1990.

\bibitem{Witten:1989vn}
E.~Witten.
\newblock Quantum field theory and the jones polynomial.
\newblock {\em Commun. Math. Phys.}, 121:351, 1989.

\bibitem{Witten:1989ys}
E.~Witten.
\newblock Gauge theories and integrable lattice models.
\newblock {\em Nucl. Phys.}, B322:629, 1989.

\bibitem{Witten:1990zr}
E.~Witten.
\newblock Gauge theories, vertex models and quantum groups.
\newblock {\em Nucl. Phys.}, B330:285, 1990.

\bibitem{Belavin:1984vg}
A.A. Belavin, A.M. Polyakov, and A.B. Zamolodchikov.
\newblock Infinite conformal symmetry in two-dimensional quantum field theory.
\newblock {\em Nucl. Phys.}, pages 333--380, 1984.

\bibitem{Moore:1989ly}
G.W. Moore and N.~Seiberg.
\newblock Classical and quantum conformal field theory.
\newblock {\em Commun. Math. Phys.}, 123:177, 1989.

\bibitem{Fradkin:1998ve}
E.H. Fradkin, C.~Nayak, A.M. Tsvelik, and F.~Wilczek.
\newblock A chern-simons effective field theory for the pfaffian quantum hall
  state.
\newblock {\em Nucl. Phys.}, B516:704, 1998.

\bibitem{:1990bh}
F.~Wilczek, editor.
\newblock {\em Fractional Statistics and Anyon Superconductivity}.
\newblock World Scientific, 1990.

\bibitem{Christensen:2007it}
N.~B. Christensen, H.~M. R{\o}nnow, D.~F. McMorrow, A.~Harrison, T.~G. Perring,
  M.~Enderle, R.~Coldea, L.~P. Regnault, and G.~Aeppli.
\newblock Quantum dynamics and entanglement of spins on a square lattice.
\newblock {\em PNAS}, 104(39):15264--15269, 2007.

\bibitem{Anderson:1950dq}
P.W. Anderson.
\newblock Antiferromagnetism. theory of superexchange interaction.
\newblock {\em Phys. Rev.}, 79:350, 1950.

\bibitem{Sawatzky:1976cr}
G.A. Sawatzky, W.~Geertsma, and C.~Haas.
\newblock Magnetic interactions and covalency effects in mainly ionic
  compounds.
\newblock {\em J. Magn. and Magnetic Mat.}, 3:37, 1976.

\bibitem{Lauchli:2006ys}
A.~Lauchli, F.~Mila, and K.~Penc.
\newblock Quadrupolar phases of the s=1 bilinear-biquadratic heisenberg model
  on the triangular lattice.
\newblock {\em Phys. Rev. Lett.}, 97:087205, 2006.

\bibitem{Schroeter:2007zr}
D.F. Schroeter, E.~Kapit, R.~Thomale, and M.~Greiter.
\newblock Spin hamiltonian for which the chiral spin liquid is the exact ground
  state.
\newblock {\em Phys. Rev. Lett.}, 99:097202, 2007.

\end{thebibliography}
\bibliographystyle{unsrt}

\end{document}